\documentclass[notspecified,article,accept,moreauthors,pdftex]{Definitions/mdpi} 

\usepackage{dsfont}
\usepackage{mymacros}

\newlength{\hcolwidth}
\firstpage{1} 
\pubvolume{1}
\issuenum{1}
\articlenumber{0}
\pubyear{2020}
\copyrightyear{2020}
\history{Received: date; Accepted: date; Published: date}

\pdfoutput=1

\Title{Detector tilt considerations in Bragg coherent diffraction imaging: a simulation study}

\Author{
	Siddharth Maddali $^{1,}$*\orcidA{}, 
	Marc Allain $^{2}$,
	Peng Li $^{2}$,
	Virginie Chamard $^{2}$,
	and Stephan O. Hruszkewycz$^{1,}$}

\AuthorNames{Siddharth Maddali, Marc Allain, Peng Li, Virginie Chamard and Stephan Hruszkewycz}

\address{%
	$^{1}$ \quad Argonne National Laboratory, 9700 S. Cass Ave, Lemont, IL 60439 (USA) \\
	$^{2}$ \quad Aix-Marseille University, CNRS, Centrale Marseille, Institute Fresnel, Marseille, France
}

\corres{Correspondence: smaddali@anl.gov}

\abstract{
This paper addresses three-dimensional signal distortion and image reconstruction issues in x-ray Bragg coherent diffraction imaging (BCDI) in the event of a general non-orthogonal orientation of the area detector with respect to the diffracted beam.
Growing interest in novel BCDI adaptations at fourth-generation synchrotron light sources has necessitated improvisations in the experimental configuration and the subsequent data analysis.
One such possibly unavoidable improvisation that is envisioned in this paper is a photon-counting area detector whose face is tilted away from the perpendicular to the Bragg-diffracted beam during acquisition of the coherent diffraction signal.
We describe a likely circumstance in which one would require such a detector configuration, along with experimental precedent at third generation synchrotrons. 
Using physically accurate diffraction simulations from synthetic scatterers in the presence of such tilted detectors, we analyze the general nature of the observed signal distortion qualitatively and quantitatively, and provide a prescription to correct for it during image reconstruction.
Our simulations and reconstructions are based on an adaptation of the known theory of BCDI sampling geometry as well as recently developed projection-based methods of wavefield propagation.
Such configurational modifications and their numerical remedies are potentially valuable in realizing unconventional coherent diffraction measurement geometries and eventually paving the way for the integration of BCDI into new materials characterization experiments at next-generation light sources.
}


\begin{document}


\section{Introduction}
\label{S:intro}
Bragg coherent x-ray diffraction imaging (BCDI) is a synchrotron-based lensless imaging technique for spatial resolution of lattice distortions on the scale of a few tens of nanometers \cite{Robinson2009,Miao2015,Hofmann2017a,Hill2018}.
BCDI is a valuable means of materials characterization owing to its ability to spatially resolve specific components of the 3D lattice strain tensor in deformed crystals, in a nondestructive manner.
This is done by reconstructing three-dimensional (3D) real-space images via phase retrieval inversion algorithms \cite{Fienup1982,Fienup1986,Zhang2016,Guizar-Sicairos2008}, 
BCDI and the related imaging technique of Bragg ptychography~\cite{Godard2011,Hruszkewycz2012,Pateras2015,Hruszkewycz2017a,Hill2018} together constitute an important set of nano-scale imaging modalities for compact as well as extended single crystal materials.

With the several orders of magnitude increase in coherent flux at fourth-generation synchrotron light sources (\emph{e.g.}, ESRF-EBS and the upcoming APS-U), coherent diffraction methods will play an increasingly important role in the 3D characterization of materials structure at the nanoscale.
The need to incorporate these techniques into existing measurement pipelines will create a requirement for flexible and unconventional diffractometer geometries for smooth functioning of these multimodal workflows.
Such a situation may arise when it is difficult to rotate a conventional detector arm into the required position to interrogate a region of interest in reciprocal space, but BCDI capabilities are nevertheless required. 

As an example, an experiment might require enhanced strain sensitivity corresponding to a higher-order Bragg reflection that may be outside the accessible rotation range of a conventional detector arm. 
In such a case, one may favor an alternate configuration such as a wall-mounted BCDI detector configuration in which the physical detector chip will not be perpendicular to the exit beam.
Another example from a recent work describes the unprecedented nanoscale strain mapping on individual crystalline domains in a poly-grain material by exploiting the partial coherence of a high-energy x-ray beam (52 keV) typically used for meso-scale orientation and strain characterization~\cite{Maddali2020b}. 
The high beam energy necessitated a sample-detector distance of $\sim 7$ m in order to resolve diffraction fringes, which was achieved with a BCDI detector mounted along the far wall and whose face was perpendicular to the downstream direction, instead of the diffracted beam (Fig.~\ref{fig:detmountings}). 
These examples indicate the possible increasing need for BCDI with unconventional detector configurations at future beamlines.

In anticipation of the increased demand for such BCDI capabilities, this paper addresses the issue of real-space image reconstruction from a BCDI signal distorted by an arbitrary detector orientation.
Specifically, we adapt the well-known theory of BCDI coordinate transformations to correctly render a distortion-free 3D image of a synthetic scatterer, from a simulated BCDI scan with a tilted detector. 
The tilt-distorted 3D diffraction pattern is obtained using a customized Fourier transform -based forward propagation model that explicitly takes into account an arbitrary detector tilt.
This method draws from the geometric theory of BCDI developed in Refs.~\cite{Maddali2020a,Li2020} and is demonstrated here with simulations.

It is worth noting that a limited form of this geometric quantification is performed by the \emph{xrayutilities} software package~\cite{Kriegner2013}. 
Specifically, it addresses the case in which the detector tilt can be decomposed as two independent tilts along the mutually perpendicular sampling directions defined by the pixels. 
Our work here generalizes this treatment to an arbitrary detector tilt relative to the perpendicular to the exit beam that need not correspond to this decomposition criterion, \emph{e.g.}, when the detector face is rotated about the exit beam.
An actual imaging experiment of this kind was performed recently~\cite{Maddali2020b}, a generalized version of which we illustrate in this paper. 

\begin{figure}
	\centering
	\includegraphics[width=0.6\hcolwidth]{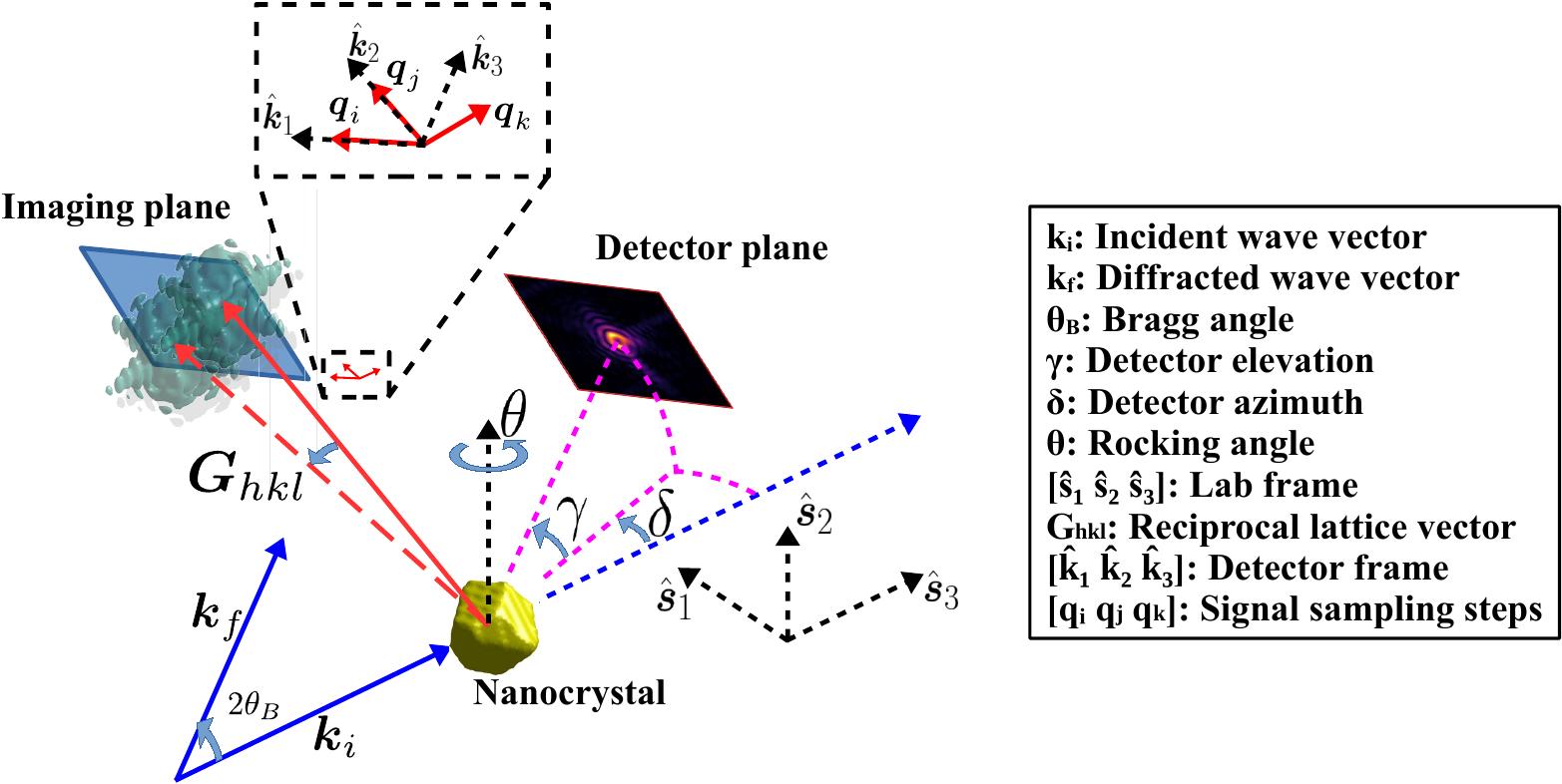}
	\caption{
		BCDI schematic with all the essential degrees of freedom and the relevant coordinate frames of reference.
	}
	\label{fig:expt}
\end{figure}

As a recap of the representative experimental schematic~\cite{Maddali2020a}, we refer to Fig.~\ref{fig:expt} that depicts the following features:
\begin{itemize}
	\item	The scattering geometry with the incident and diffracted wave vectors $\bs{k}_i$ and $\bs{k}_f$ respectively, along with the scattering angle $2\theta_B$.
			$\bs{G}_{hkl}$ is the reciprocal lattice vector corresponding to the active Bragg reflection. 
			Here, $\norm{\bs{k}_i} = \norm{\bs{k}_f} = 1/\lambda$ in the crystallographers' convention, where $\lambda$ is the wavelength of the nominally monochromatic x-rays.
			Their respective directions are given by those of the incident and diffracted beams. 
	\item	The orthonormal laboratory frame $\bs{B}_\text{lab} = [\unitvector{s}_1~\unitvector{s}_2~\unitvector{s}_3]$.
	\item	The orthonormal frame $\bs{B}_\text{img} = [\unitvector{k}_1~\unitvector{k}_2~\unitvector{k}_3]$, where $\unitvector{k}_1$ and $\unitvector{k}_2$ span the plane normal to the exit beam (hereafter referred to as the `imaging plane'). 
			The direction $\unitvector{k}_3$ is perpendicular to the imaging plane along the nominal exit beam direction. 
			This is identical to the frame $\bs{B}_\text{det}$ from Ref.~\cite{Maddali2020a}.
	\item	The rocking angle $\theta$, in this case about $\unitvector{s}_2$, but in general about any direction permissible by the experimental arrangement.
	\item	Most importantly, the discrete sampling steps in Fourier space corresponding to the detector pixels, represented as the columns of a $3 \times 3$ matrix: $\bs{B}_\text{recip} = [\bs{q}_i~\bs{q}_j~\bs{q}_k]$. 
			Here $\bs{q}_i$ and $\bs{q}_j$ are the reciprocal-space steps defined by the pixel dimensions along $\unitvector{k}_1$ and $\unitvector{k}_2$. 
			The incremental migration of the 3D diffraction signal through the imaging plane by virtue of the $\theta$ rotation is denoted in this convention by $-\bs{q}_k$ and is independent of the detector pixel dimensions.
\end{itemize}

The azimuth and elevation angles $\delta$ and $\gamma$ in Fig.~\ref{fig:expt} are specific to the configuration at the 34-ID-C end station of the Advanced Photon Source and denote typical diffractometer-based parameterizations of the detector position.
In BCDI the column vectors of $\bs{B}_\text{recip}$ are not mutually orthogonal owing to the geometry of $\bs{q}_k$ (Ref.~\cite{Maddali2020a} contains a derivation of this fact).
However the detector is typically arranged with its face normal to the exit beam such that $\bs{q}_i \parallel \unitvector{k}_1$ and $\bs{q}_j \parallel \unitvector{k}_2$, as seen in Fig.~\ref{fig:expt}.

Let the dimensions of the acquired data set be $N_1 \times N_2 \times N_3$, where $N_1$ and $N_2$ denote the pixel span of the detector and  $N_3$ the number of discrete steps in the rocking direction ($\theta$ in Fig.~\ref{fig:expt}). 
The thesis of Ref.~\cite{Maddali2020a}  is: given that $\bs{B}_\text{recip} = [\bs{q}_i~\bs{q}_j~\bs{q}_k]$ can be computed from the experimental geometry, the real-space sampling steps associated with the three axes of the digital phase-retrieved object may be similarly expressed as the columns of another matrix $\bs{B}_\text{real}$. 
Generally, the mutual non-orthogonality of the columns of $\bs{B}_\text{real}$ implies a non-orthogonality in the sampling of the reconstructed 3D object.
It has been shown~\cite{Maddali2020a} that $\bs{B}_\text{real}$  is given by:
\begin{align}
	\bs{B}_\text{real} &= \bs{B}_\text{recip}^{-T} \mathcal{D} \label{eq.thesis}\\
	\text{where } \mathcal{D} &= \left[
		\begin{matrix}
			N_1^{-1} & &  \\
			 & N_2^{-1} & \\
			 & & N_3^{-1}
		\end{matrix}
		\right] \notag
\end{align}
and $-T$ equivalently denotes the inverse of the transpose or the transpose of the inverse.

The phase-retrieved array containing the real-space scatterer, combined with knowledge of the shear encoded in $\bs{B}_\text{real}$, is sufficient for accurate, un-distorted rendering of the scatterer with one of several available software packages for 3D visualization. 
For the interested reader, the method to directly compute gradients on a grid of such sheared sample points (required to convert the scatterer's complex phase to a spatially resolved lattice strain field) is provided in the appendix of~\cite{Maddali2020a}.

\section{A tilted detector}
\label{S:tilteddetector}
\begin{figure}
	\centering
	\includegraphics[width=0.5\hcolwidth]{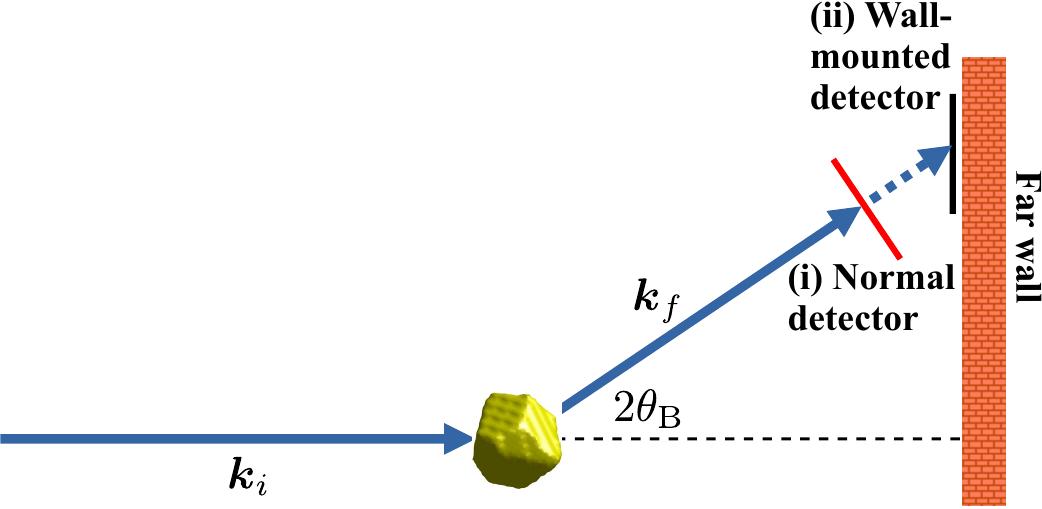}
	\caption{
		Schematic of a BCDI diffraction geometry with different detector configurations: 
		(i) the traditional mounting normal to the exit beam which is arranged by diffractometer rotation stages and a radial arm, and (ii) an unconventional wall-aligned mounting (\emph{i.e.} normal to the incident beam).
		The latter is easily achieved with inexpensive translation stages that move the detector parallel to the far wall of the experimental enclosure.
	}
	\label{fig:detmountings}
\end{figure}

The shear-correcting coordinate inversion method summarized in Section~\ref{S:intro} generalizes to any BCDI configuration provided the sampling basis matrix $\bs{B}_\text{recip}$ is properly parameterized according to the experimental degrees of freedom.
The detector plane was assumed perpendicular to the exit beam, an arrangement typically ensured in BCDI by fixing the detector on a radial arm, facing inwards and pointed directly at the mounted scatterer (case (i) in Fig.~\ref{fig:detmountings}).
In this section, we demonstrate the flexibility of the sampling basis formalism of Eq.~\eqref{eq.thesis} in addressing the general case when $\bs{q}_i$ and $\bs{q}_j$ are not aligned parallel to $\unitvector{k}_1$ and $\unitvector{k}_2$ respectively.
As demonstrated in Ref.~\cite{Maddali2020b}, such a situation may arise in the design of future BCDI facilities in which the detector configuration is dictated not by diffractometer rotations such as $\gamma$ and $\delta$, but relatively inexpensive translation stages, an example of which is shown in Fig.~\ref{fig:detmountings}, case (ii). 
Such a modification would greatly simplify the design of a BCDI measurement, with the burden of correcting for the tilt-induced signal distortion being placed on numerical methods.

We first consider a simplifying assumption.
We assume an ideal detector with pixels capable of perfect response, which faithfully register an incident photon in its entirety.
We further assume that the pixels are not susceptible to energy redistribution due to the passage of the incident radiation through multiple adjacent pixels owing to the slanted propagation path~\cite{Rueter2017}.
This undesirable feature of real-world detectors would result in a blurring effect of the acquired signal whose correction, while in principle addressable as an additional deconvolution problem, could involve details of detector chip design, thereby complicating the image reconstruction process.
Numerical corrections to address this blurring issue are detector-specific and therefore outside the scope of this work.

\begin{figure}
    \centering
    \includegraphics[width=0.5\hcolwidth]{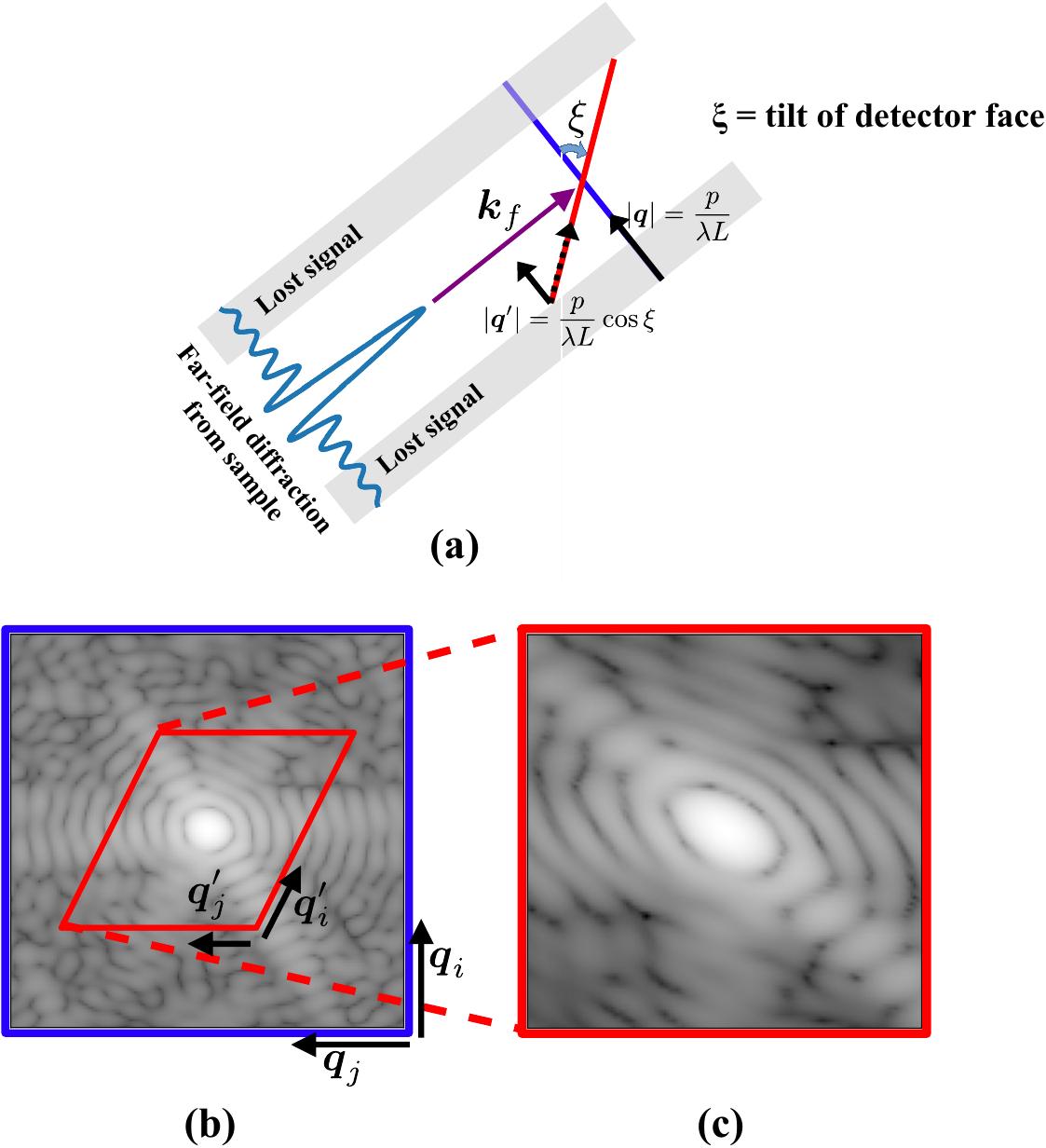}
    \caption{
		\textbf{(a)} Difference in sampling step when a 1-D line  detector is placed perpendicular to the exit beam (blue), compared to when it is tilted with respect to this orientation (red).
		The effective sampling step $\left|\bs{q}^\prime\right|$ of the finite size of the pixels in the tilted case is smaller by a factor of $\cos\xi$ than the un-tilted case, as seen by the dashed arrow.
		\textbf{(b)} A similar scenario in the case of a 2-D area detector, in which the general tilt of the detector is parameterized by not 1, but 3 angles (see~\ref{S:tiltmath} and Table~\ref{tab:tilts} for the specific parameterization and values used in this paper).  
		This compares effective 2-D signal coverage when the detector is perpendicular to the exit beam (blue), and tilted arbitrarily (red).
		\textbf{(c)} The effective signal measured therein.
		$(\bs{q}_i, \bs{q}_j)$ and $(\bs{q}_i^\prime, \bs{q}_j^\prime)$ denote the imaging plane sampling steps implied by the finite size of the area detector pixels, in the un-tilted and tilted cases.
    }
    \label{fig:tilted_detector}
\end{figure}

We now refer to the schematic in Fig.~\ref{fig:tilted_detector}.
In ~\ref{fig:tilted_detector}(a), we see a simplified one-dimensional detector arranged to capture the peak of a Bragg reflection at its center, but tilted away from the imaging plane by an angle $\xi$. 
Here we denote the distance between the object and the center of the detector as $L$ and the pixel pitch as $p$.
This arrangement renders the extent of angular space queried by the detector smaller by a factor of $\cos\xi$ (note the region of angular information `lost' to the BCDI measurement).
The Fourier space norm of the pixel step is no longer $p/\lambda L$ as shown in \cite{Maddali2020a}, but $(p/\lambda L) \cos\xi$.
The scattered intensity in the region of lost information does not contribute to the acquired signal. 
Even in the case of high signal-to-noise ratio (SNR), this would result in a measurement with missing higher-order Fourier components and therefore a reconstruction with necessarily lower spatial resolution.
A more general treatment of this one-dimensional detector case is found in Ref.~\cite{Kriegner2013}.

We wish to generalize this idea to the case of a two-dimensional area detector arbitrarily tilted with respect to the exit beam.
We refer to Fig.~\ref{fig:tilted_detector}(b), in which the diffraction signal within the blue outline depicts what would be measured if the detector face were aligned with the imaging plane (\emph{i.e.}, the plane of the figure). 
The exit beam $\bs{k}_f$ enters the imaging plane perpendicular to the figure.
Also shown is the projection of the detector face if it were tilted arbitrarily (red outline), effectively a sheared window when viewed along the exit beam. 
The region of lost signal information is now the area in between the blue and red quadrilaterals.

We first define the mutually perpendicular sampling steps $\bs{q}_i^{\text{(tilt)}}$ and $\bs{q}_j^{\text{(tilt)}}$ as fixed to the physical pixels of the area detector, and which rotate along with the detector as it is tilted away from the imaging plane.
For a normal detector, this definition of the $\bs{q}_{i,j}^{\text{(tilt)}}$ coincides with that of $\bs{q}_{i,j}$ from Ref.~\cite{Maddali2020a}.
For a tilted detector, these vectors no longer lie in the imaging plane, and we now seek their respective projections $\bs{q}_i^\prime$ and $\bs{q}_j^\prime$ in the imaging plane, akin to the situation in Fig.~\ref{fig:tilted_detector}(a).
Owing to the arbitrary detector tilt, the projections $\{\bs{q}_i^\prime,\bs{q}_j^\prime\}$ are not orthogonal in general, even though they still span the imaging plane.
The projection operator $\bs{P}$ into the imaging plane is a $3 \times 3$ matrix defined by:
\begin{equation}
	\bs{P} = \mathcal{I} - \unitvector{k}_3 \unitvector{k}_3^T
    \label{eq.projection}
\end{equation}
where $\mathcal{I}$ is the $3 \times 3$ identity matrix and $\unitvector{k}_3$ is the third axis of the detector frame (along the direction of the exit beam), treated as a $3 \times 1$ column vector. 
With this, we can compute the effective sampling steps in the imaging plane:
\begin{align}
	\bs{q}_i^\prime = \bs{P}\bs{q}_i^{\text{(tilt)}}
    \label{eq.proj_i} \\
	\bs{q}_j^\prime = \bs{P}\bs{q}_j^{\text{(tilt)}}
    \label{eq.proj_j}
\end{align}
The basis vectors $\bs{q}_i$ and $\bs{q}_j$ in the matrix expression for $\bs{B}_\text{recip}$ are respectively replaced by $\bs{q}_i^\prime$ and $\bs{q}_j^\prime$ from Eqs.~\eqref{eq.proj_i} and~\eqref{eq.proj_j}.
The modified $\bs{B}_\text{recip}$ in turn allows us to calculate the correct $\bs{B}_\text{real}$ according to Eq.~\eqref{eq.thesis}.

We note the following: 
\begin{itemize}
	\item	The third sampling vector $\bs{q}_k$ is not modified by the tilt of the detector, since it depends only on the Bragg reflection of interest and the direction of scatterer rocking.
	\item	The information of the detector tilt is introduced into $\bs{q}_i^\prime$ and $\bs{q}_j^\prime$ not through the projection operator $\bs{P}$, but the now out-of-plane vectors $\bs{q}_i^{\text{(tilt)}}$ and $\bs{q}_j^{\text{(tilt)}}$.
	\item	As seen in Fig.~\ref{fig:tilted_detector}(c), a tilted detector results in the measurements of a distorted diffraction pattern $\norm{\Psi^\prime(\bs{q})}^2$ which in turn corresponds to the distorted wave field $\Psi^\prime(\bs{q}) e^{\iota 2\pi\bs{C}_0^T\bs{q}}$.
		Here $\Psi^\prime(\bs{q})$ denotes the wave field resulting from the in-plane distortion, while the complex exponential phase ramp $e^{\iota 2\pi \bs{C}_0^T \bs{q}}$ parameterized by a constant vector $\bs{C}_0$ denotes the distribution of the phase lag in the interrogated wave field relative to the phase profile at the imaging plane. 
			This additional linearly varying  phase clearly does not influence the measured signal, and is therefore not considered from here on, assuming the measurements are in fact made in the far-field regime.
\end{itemize}

\section{Simulation results}
\label{S:simulation}
We now demonstrate the reconstruction of a synthetic digital object from simulated BCDI scans acquired at various detector tilts.
For all the following simulations, we adopt the self-conjugate detector frame $\bs{B}_\text{img} = [\unitvector{k}_1~\unitvector{k}_2~\unitvector{k}_3]$ defined earlier in which to render the original and reconstructed objects, as well as the Fourier-space signal. 
For simplicity, the synthetic object in question is a phase-less pyramid with a square base, with well-defined facets and edges.
The forward model to simulate the signal acquired using tilted detectors is described in detail in Appendix~\ref{S:tiltmath}, along with a summary of the various detector tilts used in the simulations.
These manipulations are predicated upon the projection-based far-field propagation method, whose detailed derivation is the subject of Ref.~\cite{Li2020}.
Here, in the interests of highlighting the detector tilt-induced object shear and its correction, we bypass the phase retrieval process altogether and merely obtain the `reconstructions' of the scatterer from the inverse FFT of the simulated wave fields, and compare their morphologies before and after the distortions that arise from Eqs.~\eqref{eq.proj_i} and~\eqref{eq.proj_j} have been corrected.
Under these circumstances, the inverse FFT is simply a proxy for the phase retrieval solution in the limit of infinite signal to noise ratio.
\begin{figure}
	\centering
	\includegraphics[width=\hcolwidth]{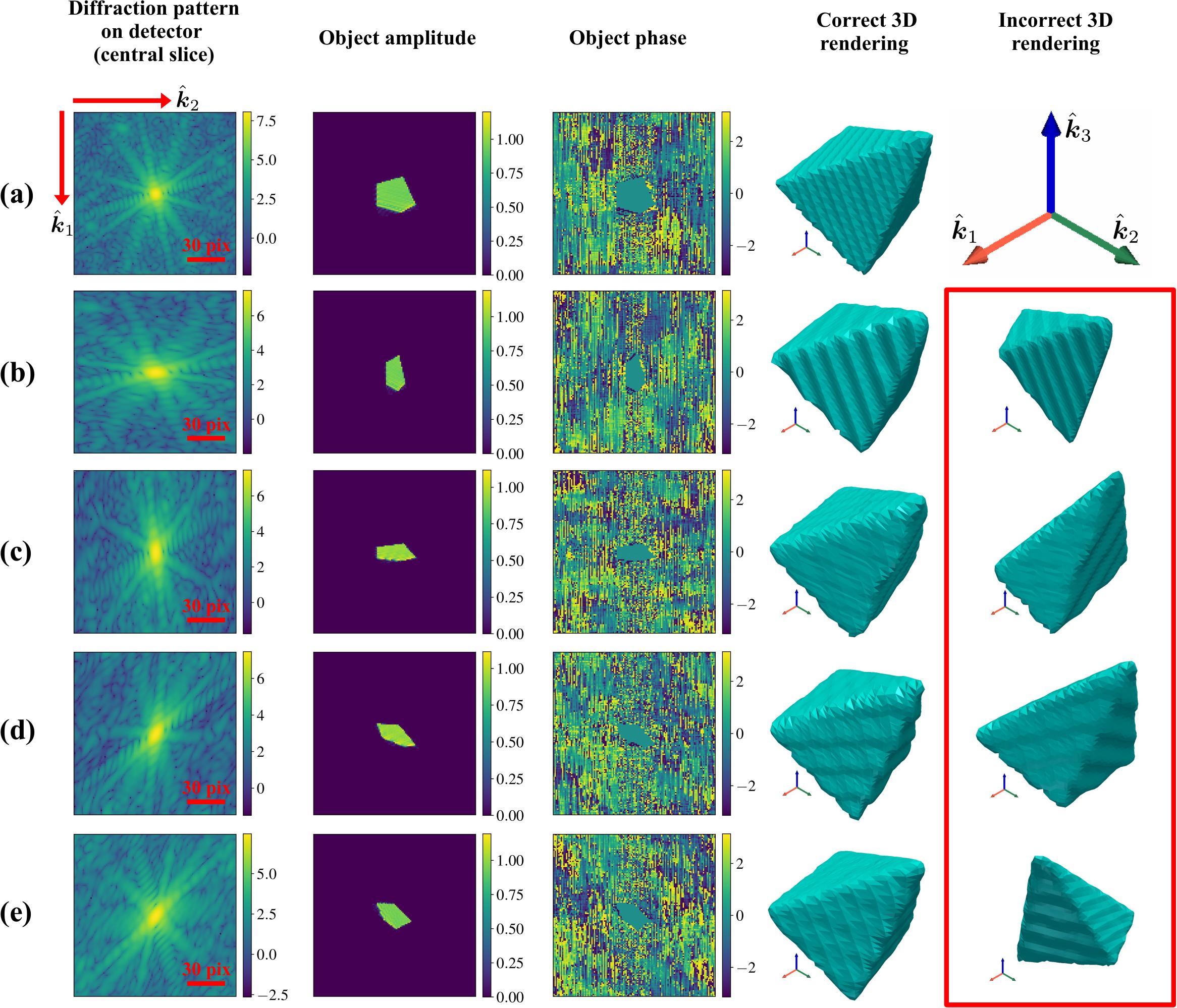}
	\caption{
		Simulations of the signal distortion and object reconstructions, for different detector tilts.
		\textbf{(a)}
			The no-tilt case of conventional BCDI, in which the detector is normal to the exit beam. 
			The first image shows the central slice of the measured 3D diffraction pattern.
			The second and third images are the central amplitude and phase cross-sections of the `reconstructed' object (\emph{i.e.}, a simple inverse FFT).
			The fourth image is the inferred object isosurface corresponding to the tilt-corrected sampling basis $\bs{B}_\text{real}$ (see Ref.~\cite{Maddali2020a}).
		\textbf{(b)---(e)} The corresponding images for various non-zero detector tilts parameterized by the angles $\xi$, $\zeta$ and $\phi$ (see \ref{S:tiltmath} and Table~\ref{tab:tilts} for definitions and specific values).
		In each of the (b)---(e) subfigures, the last image is the inferred object if the detector tilt is not taken into account, and it is na\"{i}vely assumed that the detector is perpendicular to the exit beam, as in conventional BCDI.
		In each of the diffraction patterns (first column), the red bar indicates a span of 30 pixels.
}
	\label{fig:tiltresults}
\end{figure}

Fig.~\ref{fig:tiltresults} shows the simulated diffraction signals of the phase-less pyramid than has been arbitrarily oriented in the detector frame, when the detector is tilted in different ways with respect to the exit beam.
We note the `stretched' nature of the diffraction patterns in Figs.~\ref{fig:tiltresults}(b), \ref{fig:tiltresults}(c), \ref{fig:tiltresults}(d) and \ref{fig:tiltresults}(e) along various directions owing to the tilted detector, when compared to the diffraction pattern in \ref{fig:tiltresults}(a) (corresponding to perpendicular detector).
For example, \ref{fig:tiltresults}(b) shows the diffraction when the detector is tilted by $60^\circ$ about $\unitvector{k}_1$, which appears like the diffraction pattern in \ref{fig:tiltresults}(a), but stretched along $\unitvector{k}_2$. 
This signal contains corrupted high-frequency fringe information along this direction, which in a real-world measurement would translate to deficiency of spatial resolution that will manifest in the blurred edges of the reconstructed object.
Similarly, Fig.~\ref{fig:tiltresults}(c) shows the signal when the detector is rotated by $60^\circ$ about $\unitvector{k}_2$ (\emph{i.e.}, $\xi = 60^\circ$, $\zeta = 90^\circ$), resulting in a stretch along $\unitvector{k}_1$ and a corresponding blurring along the edges of the recovered object.
Figs.~\ref{fig:tiltresults} also show the distorted diffraction signal in the case of more complicated detector tilts (see \ref{S:tiltmath} for a full summary).
In each row, the second and third images show the amplitude and phase cross sections of the recovered object with a simple inverse FFT of the diffracted wave field.
This is what would have been recovered in an actual phase retrieval reconstruction.

In each case, the tilt-corrected pyramid isosurface is shown along with the corresponding isosurface when one disregards the detector tilt (fourth and fifth images in each row respectively). 
More specifically, the fourth image in each row corresponds to the correct real-space sampling basis $\bs{B}_\text{real}$, obtained from Eq.~\eqref{eq.thesis} after properly accounting for the detector tilt  (the tilt corrections in each case being given by Eqs.~\eqref{eq.proj_i} and \eqref{eq.proj_j}). 
The fifth image in each row of Fig.~\ref{fig:tiltresults} corresponds to the inferred object without accounting for the detector tilt.
Clearly, the tilt-corrected isosurface agrees with the original pyramid in Fig.~\ref{fig:tiltresults}(a) in terms of morphology and orientation.

It was observed that phase variation in the interior of the (nominally phase-less) recovered object following the forward propagation was insignificant, $\sim 10^{-4}$ radians (see the Jupyter notebook in the Supplementary Material).
Further, we note that Fig.~\ref{fig:tiltresults}(e) corresponds to the area detector being rotated by $73^\circ$ about the exit beam direction $\unitvector{k}_3$. 
In contrast, the \emph{xrayutilities} formalism assumes a conventional `3S+2D' goniometer configuration (3 sample rotations, 2 detector rotations with the detector nominally fixed perpendicular to the exit beam), and corrects for small rotations of the detector about the exit beam direction.
In this sense, the rotation of $73^\circ$ is outside the scope of analysis by \emph{xrayutilities}.

The striations on the pyramid faces in panels (b), (c), (d) and (e) can also be explained by the corruption of the high-frequency Fourier components caused by cyclic aliasing in the simulation process.
In a real-world measurement with a tilted detector, the high-frequency Fourier components are not aliased, but genuinely lost to the measurement because they fall outside the aperture defined by the projected area of the detector (see Fig.~\ref{fig:tilted_detector}(b)).
In general this results in a loss of spatial resolution in the reconstructed object.

\section{Conclusion}
\label{S:conclusion}
We have derived a geometric correction for the morphology of a reconstructed scatterer in a BCDI measurement with a detector tilted with respect to the diffracted beam.
The correction method demonstrated is seen to be a straightforward generalization of the mathematial theory developed in Ref.~\cite{Maddali2020a}. 
We have successfully validated our theory by developing a customized forward model of the distorted diffraction signal acquired by a tilted detector and adapting the coordinate transform theory from Ref.~\cite{Maddali2020a} to obtain the correct post- phase retrieval 3D rendering of the original object. 
Thereby we have demonstrated potential flexibility in physical BCDI experiment design by offloading the computational burden of inverting a distorted signal to numerical methods.

\authorcontributions{
	Conceptualization, methodology, validation, S.M., M.A. and P.L.; 
	writing -- review and editing, S.M., M.A., V.C. and S.O.H.; visualization, S.M.; 
	project administration, S.O.H.; funding acquisition, S.O.H. 
	All authors have read and agreed to the published version of the manuscript.
}

\funding{
The conceptualization of the tilted-detector BCDI measurement and development of the required wave propagators  were supported by the US Department of Energy (DOE), Office of Science, Basic Energy Sciences, Materials Science and Engineering Division. 
The generalized, geometry-aware propagator theory underlying this effort was developed with the support of the European Research Council (European Union’s Horizon H2020 research and innovation program grant agreement No. 724881). 
}

\conflictsofinterest{The authors declare no conflicts of interest that would affect the publication of this article in \emph{Crystals}.}

\abbreviations{The following abbreviations are used in this manuscript:\\
\noindent 
\begin{tabular}{@{}ll}
	BCDI & Bragg coherent diffraction imaging \\
	ESRF-EBS & European Synchrotron Radiation Facility - Extremely Brilliant Source \\
	APS-U & Advanced Photon Source - Upgraded \\
	FFT & Fast Fourier transform \\
	IFFT & Inverse Fourier transform \\
	SNR & Signal-noise ratio
\end{tabular}}

\appendixtitles{yes}
\appendix
\section{Simulating diffraction with a tilted detector through Fourier space resampling}
\label{S:tiltmath}
Consider a compact crystalline scatterer denoted by the complex scalar field $\psi(\bs{r})\equiv \psi(x, y, z)$ whose coordinates are defined in the orthonormal detector frame $\bs{B}_\text{img} = [\unitvector{k}_1~\unitvector{k}_2~\unitvector{k}_3]$.
In a BCDI measurement, the squared modulus of its Fourier transform is measured slice by slice using an area detector whose Fourier-space imaging plane is displaced by integer multiples of $\bs{q}_k$, defined by a single step along the rocking curve. 
Let $\bs{r} = [ x~y~z]^T$ and $\bs{q} = [q_x~q_y~q_z]^T$ be conjugate spatial coordinates corresponding to the object wave and Bragg-diffracted far-field wave respectively.
Further, if $\bs{q}_k \equiv [q_k^{(1)}~~q_k^{(2)}~~q_k^{(3)}]^T$ in the detector frame, then the $n$th slice of the diffracted wave field is derived from the projection-slice theorem~\cite{Bracewell1956,Bracewell1990} and reads akin to Eq. (34) in Ref.~\cite{Li2020}:
\begin{equation}\label{eq.fft_spthm}
	\underbrace{\Psi_n\left(q_x, q_y\right)}_\text{slice of 3D Fourier transform} =  \underbrace{
	\int_\mathds{R}dx \int_\mathds{R}dy~
	e^{-\iota 2\pi \left[x\left(q_x + nq_k^{(1)}\right) + y\left(q_y + nq_k^{(2)}\right)\right]}
	}_\text{2D Fourier transform}
	\underbrace{
		\int_\mathds{R}dz~ e^{-\iota 2\pi z nq_k^{(3)}}
	}_\text{projection}
	\psi(x, y, z)
\end{equation}
\emph{i.e.} the $n$th slice of the scattered 3D wave field whose intensity is accessed by the area detector is equal to the 2D Fourier transform of the modulated projection of the scatterer, evaluated at the 2D points $(q_x + nq_k^{(1)}, q_y + nq_k^{(2)})$.
The modulation in question is the phase factor $e^{-i 2\pi znq_k^{(3)}}$.
The expression~\eqref{eq.fft_spthm} is evaluated numerically by means of the two-dimensional Fourier transform operator $\mathcal{F}_\text{2D}$ and the projection operator $\mathcal{R}_3$ along the $\unitvector{k}_3$-direction by:
\begin{equation}\label{eq.fft_spthm_op}
	\Psi_n(q_x, q_y) = \mathcal{F}_\text{2D} \mathcal{R}_3 e^{-\iota 2\pi \bs{r}^T n\bs{q}_k} \psi(\bs{r})
\end{equation}
One may rewrite Eq.~\eqref{eq.fft_spthm} more explicitly in terms of the two-dimensional quantities $\bs{r}_\text{2D} \equiv [x~y]^T$, $\bs{q}_\text{2D} \equiv [q_x~q_y]^T$ and $\bs{q}_{k,\text{2D}} \equiv [q_k^{(1)}~q_k^{(2)}]^T$:
\begin{equation}\label{eq.fft_spthm_compact}
	\Psi(\bs{q}_\text{2D}) = 
	\int_\mathds{R}dx \int_\mathds{R}dy~
	e^{-\iota 2\pi \bs{r}_\text{2D}^T \left(\bs{q}_\text{2D} + n\bs{q}_{k,\text{2D}}\right)}
	\int_\mathds{R}dz~ e^{-\iota 2\pi z n q_k^{(3)}}
	\psi(\bs{r}_\text{2D}, z)
\end{equation}

\begin{figure}
	\centering
	\includegraphics[width=\hcolwidth]{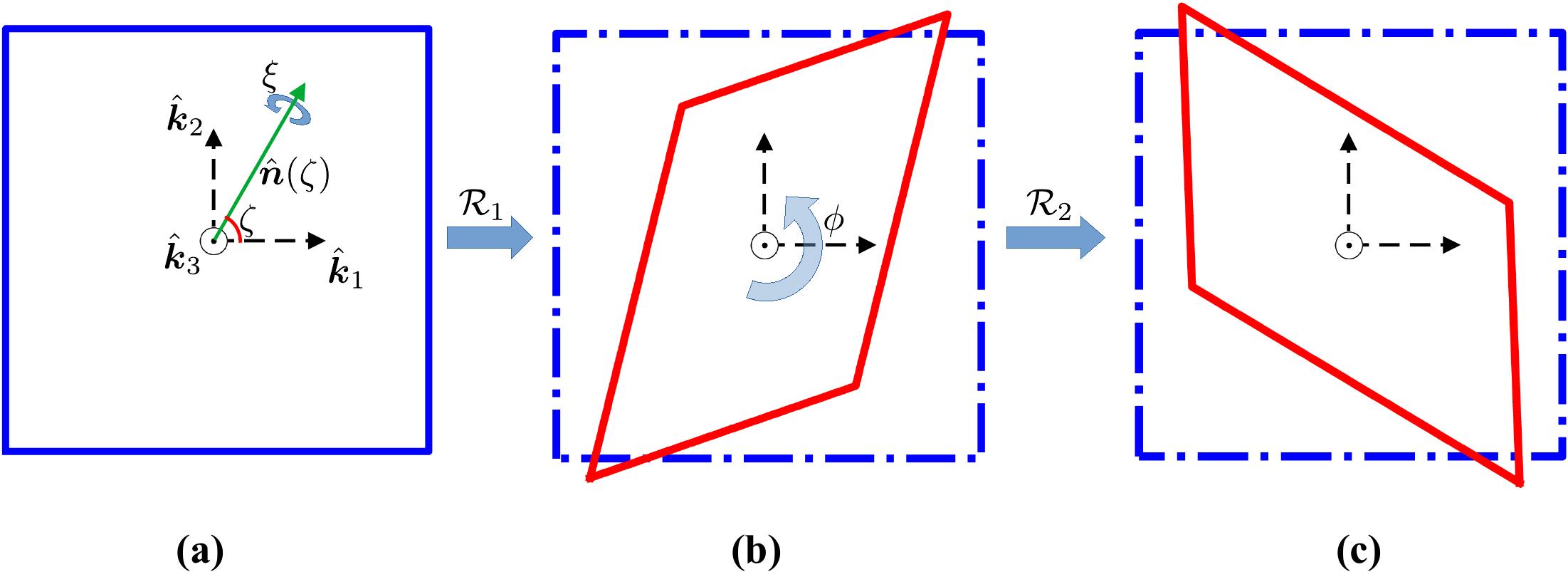}
	\caption{
		Sequence of transformations that parameterize the effective detector tilts in this paper. 
		In each sub-figure, the plane of the figure is the imaging plane, with the orthonormal basis $\bs{B}_\text{img}$ shown.
		$\unitvector{k}_3$ exits the plane of the figure.
		The blue outline shows delineates the projected area of the un-tilted detector when viewed from this vantage point, while the red area delineates the projected area of the tilted detector.
		\textbf{(a)} The initially un-tilted detector, 
		\textbf{(b)} the projected detector area (outlined in red) after the transformation $\mathcal{R}_1$ which rotates it by an angle $\xi$, about the imaging-plane axis $\unitvector{n}(\zeta) \equiv \left[\cos\zeta~\sin\zeta~0\right]^T$, 
		\textbf{(c)} the (final) projected area of the tilted detector after the subsequent rotation $\mathcal{R}_2$ by an angle $\phi$ about $\unitvector{k}_3$.
		The specific case shown here corresponds to $\xi = 60^\circ$, $\zeta = 60^\circ$, $\phi = 73^\circ$ (\emph{i.e.}, the last row in Table~\ref{tab:tilts}).
	}
	\label{fig:tiltdemo}
\end{figure}

In order to model the tilt of the detector face, we employ the axis-angle parameterization of a rotation matrix $\mathcal{R}(\alpha, \unitvector{n})$ (described in \cite{Maddali2020a} Eq.~(19)), acting on the columns of the $3 \times 2$ matrix $[ \bs{q}_i~\bs{q}_j]$. 
We recall that these columns represent the pixel steps in perpendicular directions along the face of the detector.
The tilt is modeled by the following two rotations applied in order:
\begin{enumerate}
	\item	A rotation $\mathcal{R}_1 = \mathcal{R}(\xi, \unitvector{n}(\zeta))$ by an angle $\xi$ about an axis $\unitvector{n}(\zeta) \equiv \unitvector{k}_1 \cos\zeta + \unitvector{k}_2 \sin\zeta = [\cos\zeta~\sin\zeta~0]^T$ in the $(\unitvector{k}_1, \unitvector{k}_2)$ imaging plane followed by\ldots
	\item	A rotation $\mathcal{R}_2 = \mathcal{R}(\phi, \unitvector{k}_3)$ by an angle $\phi$ about the exit beam direction.
\end{enumerate}
This sequence of detector rotations constitutes an effective departure of the detector from the normal position, parameterized by $(\zeta, \xi, \phi)$. 
These tilts are illustrated in Fig.~\ref{fig:tiltdemo}. 

In short, if we define $\mathcal{R}_2\mathcal{R}_1 \equiv \mathcal{R}_\text{tilt}(\xi, \zeta, \phi)$, then the pixel sampling steps (originally aligned along $\unitvector{k}_1$ and $\unitvector{k}_2$) are transformed due to a tilted detector in the following manner:
\begin{equation}
	\left[\bs{q}_i~\bs{q}_j\right] 
	\xrightarrow[\text{is tilted}]{\text{detector}}
	\underbrace{
		\mathcal{R}(\phi, \unitvector{k}_3)
		\mathcal{R}(\xi, \unitvector{n}(\zeta))
	}_{\mathcal{R}_2 \mathcal{R}_1}
	\left[\bs{q}_i~\bs{q}_j\right]
	 = \mathcal{R}_\text{tilt}(\xi, \zeta, \phi) \left[\bs{q}_i~\bs{q}_j\right]
\end{equation}
The in-plane sampling vectors described in Eq.~\eqref{eq.proj_i} and~\eqref{eq.proj_j} are obtained by:
\begin{equation}\label{eq.proj_ij}
	\left[\bs{q}_i^\prime~\bs{q}_j^\prime\right] = 
	\bs{P} 
	\mathcal{R}_\text{tilt}
	\left[\bs{q}_i~\bs{q}_j\right]
\end{equation}
where $\bs{P}$ is the projection operator from Eq.~\eqref{eq.projection}.
The effective tilt angle of the detector away from the normal, \emph{i.e.}, $\theta_\text{tilt}$, is in general different from the parameters $\xi$, $\zeta$ and $\phi$ and is given by:
\begin{equation}\label{eq.tiltangle}
	\cos\theta_\text{tilt} = \frac{1}{2}\left[\Tr{\mathcal{R}_\text{tilt}} - 1\right]
\end{equation}

where $\Tr{\cdot}$ denotes the matrix trace. 
As mentioned in Section~\ref{S:tilteddetector}, $\bs{q}_i^\prime$ and $\bs{q}_j^\prime$ are no longer mutually perpendicular, even though they span the imaging plane.
Of course, we ignore the extreme tilt of $\xi = 90^\circ$, in which case $\bs{q}_i^\prime \parallel \bs{q}_j^\prime$ and they no longer span the imaging plane.

We note in passing that we have expressed the tilt of the detector as a general rotation matrix, a quantity known to require 3 parameters to be unambiguously specified. 
In our case, these parameters are: (1) the in-plane orientation $\zeta$ of the first rotation axis $\unitvector{n}(\zeta)$, (2) the angle of rotation $\xi$ about this axis, and (3) the angle of rotation $\phi$ about the $\unitvector{k}_3$ direction.
The \emph{xrayutilities} library~\cite{Kriegner2013}, on the other hand, restricts itself to a two-parameter tilt of the detector about mutually perpendicular directions and explicitly stops short of a full parameterization. 
In this sense, the formalism being developed here is more general and capable of addressing the experimental configurations beyond the scope of \emph{xrayutilities} (of which Ref.~\cite{Maddali2020b} describes an instance).

We reiterate that there is no effect of the tilted detector on the third sampling vector $\bs{q}_k$, which is determined solely by the manner of rotation of the scatterer (`rocking') during the measurement.
We next define the projection operator $\bs{K} \equiv [\unitvector{k}_1~\unitvector{k}_2]^T$ that extracts the first two components of its 3D vector operand \emph{i.e.} for any detector-frame 3D vector $[x~y~z]^T$, we have $\bs{K} [x~y~z]^T = [x~y]^T$.

We now seek the two-dimensional shearing operation that distorts the wave field in the imaging plane due to the detector tilt, in the manner described in Section~\ref{S:tilteddetector}.
Put differently, we seek the $2 \times 2$ shear matrix $\bs{S}$ that satisfies the following condition:
\begin{equation}
	\bs{K} \bs{P} 
	\mathcal{R}_\text{tilt}
	[\bs{q}_i~\bs{q}_j] =
	\bs{S} \bs{K} \bs{P} [\bs{q}_i~\bs{q}_j]
\end{equation}
which gives us the formal expression for the two-dimensional in-plane distortion operator:
\begin{equation}\label{eq.inplane_shear_operator}
	\bs{S} = 
	\bs{K} \bs{P} 
	\mathcal{R}_\text{tilt}[\bs{q}_i~\bs{q}_j]
	\left(
		\bs{K} \bs{P}[\bs{q}_i~\bs{q}_j]
	\right)^{-1}
\end{equation}
Thus, from Eq.~\eqref{eq.inplane_shear_operator} we are now able to determine the 2D sample points $\bs{q}_\text{2D}^\text{(t)}$ accessed by the tilted detector (superscript `\emph{t}' stands for \emph{tilt}), in terms of the sample points $\bs{q}_\text{2D}$ if the detector were not tilted:
\begin{equation}\label{eq.inplane_shear}
	\bs{q}_\text{2D}^\text{(t)} = \bs{S} \bs{q}_\text{2D}
\end{equation}
We have from Eq.~\eqref{eq.fft_spthm_compact}: 
\begin{align}
	\Psi_n\left(\bs{q}_\text{2D}^\text{(t)}\right) &= 
	\int_\mathbb{R}dx\int_\mathbb{R}dy~
	e^{-\iota 2\pi \bs{r}_\text{2D}^T \left(\bs{q}_\text{2D}^\text{(t)} + n\bs{q}_{k,\text{2D}}\right)}
	\int_\mathbb{R}dz~e^{-\iota 2\pi zn q_k^{(3)}}
	\psi(\bs{r}_\text{2D}, z) \label{eq.k2d_t_instead} \\
	&= \int_\mathbb{R}dx\int_\mathbb{R}dy~
		e^{
			-\iota 2\pi 
			\left(\bs{S}^T \bs{r}_\text{2D}\right)^T
			\left(\bs{q}_\text{2D} + n\bs{S}^{-1}\bs{q}_{k,\text{2D}}\right)
		}
		\int_\mathbb{R} dz~e^{-\iota 2\pi znq_k^{(3)}}
		\psi(\bs{r}_\text{2D}, z)	\label{eq.firstsub}
\end{align}
A change of integration variable $\bs{r}_\text{2D} \longrightarrow \tilde{\bs{r}}_\text{2D} \equiv \bs{S}^T\bs{r}_\text{2D}$ in Eq.~\eqref{eq.firstsub} gives: 
\begin{equation}
	\Psi_n\left(\bs{q}_\text{2D}^\text{(t)}\right) = 
		\frac{1}{\det(\bs{S})}
		\int_\mathbb{R}\tilde{dx} \int_\mathbb{R}\tilde{dy}
		e^{
			-\iota 2\pi \tilde{\bs{r}}_\text{2D}^T 
			\left(\bs{q}_\text{2D} + n\bs{S}^{-1}\bs{q}_{k,\text{2D}}\right) 
		}
		\int_\mathbb{R}dz~e^{-\iota 2\pi nzq_k^{(3)}} \psi\left(\bs{S}^{-T}\tilde{\bs{r}}_\text{2D}, z\right)
		\label{eq.fft_spthm_compact_effective}
\end{equation}

We note that up to the multiplicative term $1/\det(\bs{S})$, the expression~\eqref{eq.fft_spthm_compact_effective} is completely analogous to Eq.~\eqref{eq.fft_spthm_compact}, whose operator version is Eq.~\eqref{eq.fft_spthm_op}. 
Eq.~\eqref{eq.fft_spthm_compact_effective} tells us that the far-field coherent diffraction can in fact be simulated on an arbitrarily tilted detector in a computationally efficient manner using Eq.~\eqref{eq.fft_spthm_op}, provided the following modifications are made:
\begin{enumerate}
	\item	The signal sampling shear $\bs{S}$ in the imaging plane is computed using Eq.~\eqref{eq.inplane_shear_operator}.
	\item	The Fourier-space incremental step $\bs{q}_k$ due to sample rocking is sheared in its first two dimensions by:
		$ \bs{q}_{k,\text{2D}} \longrightarrow \bs{S}^{-1} \bs{q}_{k,\text{2D}}$.
	\item	The scatterer $\psi$ is re-sampled in its first two dimensions by:
			$\psi\left(\bs{r}_\text{2D}, z\right) \longrightarrow \psi\left(\bs{S}^{-T} \bs{r}_\text{2D}, z\right)$
\end{enumerate}

This is the method adopted to obtain the 3D wave fields and subsequently the diffraction patterns incident upon a tilted detector. 
We note that for the purposes of this demonstration, we are able to generate the resampled pyramid $\psi(\bs{S}^{-T} \bs{r}_\text{2D}, z)$ analytically with relative ease from knowledge of its facet locations and orientations (see Jupyter notebook \texttt{simulatedDiffraction.ipynb} in the Supplementary Material). 
Further, even with the 3D array of a given arbitrary complex-valued scatterer that cannot be obtained by analytic functions, resampling is possible and quite readily achieved in a systematic and generalized manner using known Fourier-based methods~\cite{Unser1995,Larkin1997,Chen2000,Thevenaz2000}.

\begin{table}
	\centering
	\caption{Experimental parameters of the simulated BCDI forward model. Refer to Fig.~\ref{fig:expt} for the experimental geometry.}    
	\label{tab:experiment}
    \begin{tabular}{|c|c|c|}
    	\hline
	    \textbf{Parameter} & \textbf{Value} & \textbf{Description} \\ \hline
	    $E$ & $9$ keV & Beam energy \\ \hline
	    $\lambda$ & $1.378$ \AA & Wavelength \\ \hline
	    $\Delta\theta$ & $0.01^\circ$ & Angular increment \\ \hline
	    $L$ & $0.65$ m & Object-detector distance \\ \hline
		$\gamma$ & $12.0^\circ$ & Detector alignment (elevation) \\ \hline
		$\delta$ & $32.1^\circ$ & Detector alignment (azimuth) \\ \hline
	    $p$ & $55 \times 10^{-6}$ m & Pixel size \\ \hline
	    $(N_1,N_2,N_3)$ & $(128,128,128)$ & Pixel array dimensions \\ \hline
    \end{tabular}
\end{table}

\begin{table}[htbp]
	\centering
	\begin{tabular}{|c|c|c|c|c|}
		\hline
		$\bs{\xi}(^\circ)$ & $\bs{\zeta}(^\circ)$ & $\bs{\phi}(^\circ)$ & $\theta_\text{tilt}(^\circ)$ & Sub-figure in Fig.~\ref{fig:tiltresults} \\ \hline
		60 & 0 & 0 & 60 &  \textbf{(b)}	\\ \hline
		60 & 90  & 0 & 60 & \textbf{(c)}	\\ \hline
		60 & 60 & 0 & 60 & \textbf{(d)} 	\\ \hline
		60 & 60 & 73 & 91.76 & \textbf{(e)}	\\ \hline
	\end{tabular}
	\caption{
		Various tilt parameters for the results shown in Fig.~\ref{fig:tiltresults}.
	}
	\label{tab:tilts}
\end{table}
The BCDI forward model was simulated with a set of fixed experimental parameters, shown in Table~\ref{tab:experiment}.

\reftitle{References}


\end{document}